\documentclass[pra,aps,superscriptaddress,twocolumn,showpacs,nofootinbib]{revtex4-1}

\usepackage{placeins}
\usepackage{amsfonts, amsmath,amssymb,amsthm,bm}
\usepackage{mathrsfs}
\usepackage{graphicx}
\usepackage[usenames,dvipsnames]{color}
\usepackage{palatino}
\usepackage{fancyhdr}
\usepackage{mathrsfs}
\usepackage{multirow}
\usepackage{hyperref}

\usepackage{etoolbox}

\hypersetup{
   colorlinks=true,
   linkcolor=blue,
   citecolor=blue,
   urlcolor=Violet
}

\newcommand{\ket}[1]{\ensuremath{\left|#1\right\rangle}}

\newcommand{\ketbra}[2]{\ensuremath{\left|#1\right\rangle\!\left\langle#2\right|}}

\newcommand{\tr}[2]{\mathrm{Tr}_{#1}\left[ #2 \right]}

\renewcommand{\v}[1]{\ensuremath{\boldsymbol #1}}

\newcommand{\R}{\mathbb{R}}
\newcommand{\be}{\begin{equation}}
\newcommand{\ee}{\end{equation}}

\newcommand{\HH}{\mathcal{H}}
\newcommand{\dd}{{\bf d}}

\theoremstyle{plain}
\newtheorem{thm}{Theorem}

\theoremstyle{definition}
\newtheorem{defn}{Definition}
\theoremstyle{remark}

\begin{document}
\title{Stochastic independence as a resource in small-scale thermodynamics}
\author{Matteo Lostaglio\footnotemark}
\affiliation{Department of Physics, Imperial College London, London SW7 2AZ, United Kingdom}
\author{Markus P. M\"uller\footnotemark[2]}
\affiliation{Department of Applied Mathematics, Department of Philosophy, University of Western Ontario, London, ON N6A 5BY, Canada}
\affiliation{Perimeter Institute for Theoretical Physics, Waterloo, ON N2L 2Y5, Canada}
\affiliation{Institut f\"ur Theoretische Physik, Universit\"at Heidelberg, Philosophenweg 19, D-69120 Heidelberg, Germany}
\author{Michele Pastena\footnotemark[2]}
\affiliation{Institut f\"ur Theoretische Physik, Universit\"at Heidelberg, Philosophenweg 19, D-69120 Heidelberg, Germany}

\begin{abstract}
It is well-known in thermodynamics that the creation of correlations costs work. It seems then a truism that if a thermodynamic transformation $A \rightarrow B$ is impossible, so will be any transformation that in sending $A$ to $B$ also correlates among them some auxiliary systems $C$. Surprisingly, we show that this is not the case for non-equilibrium thermodynamics of microscopic systems. On the contrary, the creation of correlations greatly extends the set of accessible states, to the point that we can perform on individual systems and in a single shot any transformation that would otherwise be possible only if the number of systems involved was very large. We also show that one only ever needs to create a vanishingly small amount of correlations (as measured by mutual information) among a small number of auxiliary systems (never more than three). The many, severe constraints of microscopic thermodynamics are reduced to the sole requirement that the non-equilibrium free energy decreases in the transformation. This shows that, in principle, reliable extraction of work equal to the free energy of a system can be performed by microscopic engines.    
\end{abstract}

\date{September 16, 2015}

\maketitle

Single-shot thermodynamics studies non-equilibrium transformations of a small number of microscopic systems in contact with a heat bath. It departs substantially from the familiar description of equilibrium situations: the work necessary to create a state does not coincide with the work that can be extracted from it \cite{horodecki2013fundamental}; necessary and sufficient conditions for the existence of a thermodynamic transformation connecting two non-equilibrium states involve an infinite family of free energies $\{F_{\alpha}\}$ \cite{brandao2013second}; the quality of the extracted work must be carefully assessed due to fluctuations \cite{dahlsten2011inadequacy, aberg2013truly}. New tools and concepts are indeed needed in this regime, and we can now ask (and partially answer) many questions beyond those allowed in standard approaches \cite{ruch1978mixing, janzing2000thermodynamic, egloff2014measure, gour2013resource, narasimhachar2014low, halpern2014unification, cwiklinski2014towards, aberg2014catalytic, alhambra2015what, woods2015maximum}.

\begin{figure}[t!]
\includegraphics[width=1\columnwidth]{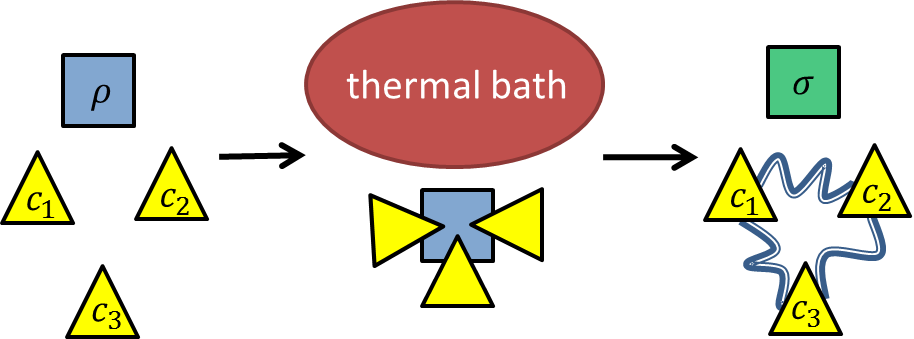}
\caption{The general scenario: a quantum state $\rho$ is transformed into a state $\sigma$ exploiting a thermal bath and auxiliary systems that ease the transformation, but are given back unchanged at the end. If correlations can be created among the auxiliary systems, we prove that such a transformation is possible if and only if $F(\rho)\geq F(\sigma)$.}
\label{fig:correlations}
\end{figure}

In this paper we focus on the role of correlations in this regime. We consider the general scenario in which, given a system in any out-of-equilibrium state $\rho$, we want to obtain a target state $\sigma$. We can use a thermal bath and auxiliary systems $c_1,\ldots,c_N$ that catalyze the transformation, but are given back unchanged. Severe constraints need to be met for such a transformation to exist \cite{brandao2013second}. We study here what happens if we allow the auxiliary systems to get correlated in the process (see Fig.~\ref{fig:correlations}).

At first glance it seems that this cannot be of any help, because the creation of correlations increases the free energy of the auxiliary systems. Hence, the argument goes, the creation of correlations is yet another obstacle to the requirement that the free energy has to decrease in the process. However, there is much more to single-shot thermodynamics than just ``the'' free energy. Surprisingly, we show that the creation of correlations greatly enlarges the set of states that can be obtained from $\rho$. Indeed, any transformation that decreases the free energy becomes possible in the single-shot regime. In other words, all transformations that would be possible in the thermodynamic limit of processing $n\rightarrow \infty$ uncorrelated copies of a system \cite{brandao2011resource, halpern2014unification} become possible on individual systems. This gives a single-shot operational meaning to the free energy and shows that if an engine can access uncorrelated auxiliary systems, it can operate as if it was reversible, even in extreme thermodynamic regimes. We also show that the correlations that one needs to create for this purpose are always vanishingly small.

In this work we will focus on quantum states initially incoherent in energy. Notice that since all states in this paper are block-diagonal in the energy eigenbasis, we identify quantum states and the vector of their eigenvalues. A new framework for incorporating the role of quantum coherence in thermodynamics, based on symmetry principles, has been put forward in \cite{lostaglio2015description, lostaglio2015quantum}. We leave for future research the question of how to unify the symmetry analysis on coherence with the present considerations on correlations.

\subsection{Severe constraints of single-shot thermodynamics}

To better understand the issues at hand before presenting our general result, we consider a paradigmatic example. Suppose we are given a qubit system $\rho$, with Hamiltonian $\mathcal{H}_S=E\ketbra{1}{1}$ and population $p$ in the ground state. To extract work from it we are allowed to bring in a thermal bath in a Gibbs state (with arbitrary Hamiltonian and fixed temperature $T$) and couple it to the system by any energy-preserving interaction \cite{janzing2000thermodynamic, brandao2011resource, horodecki2013fundamental}. We also allow for the use of catalysts, i.e.\ auxiliary systems that facilitate the transformation but are given back at the end unchanged and uncorrelated with all other systems. These transformations are called \emph{catalytic thermal operations} (see \cite{brandao2013second} and Supplemental Material A). Following \cite{horodecki2013fundamental, brandao2013second}, work extraction is explicitly modeled by introducing a ``work bit'' \cite{horodecki2013fundamental}, i.e.\ a two-level system with Hamiltonian $\mathcal{H}_W=w\ketbra{1}{1}$ that is initially in the ground state $\ket{0}$ and at the end of the transformation is found with high probability in the excited state $\ket{1}$:

\be
\label{example}
\rho \otimes \ketbra{0}{0} \longrightarrow \gamma_S \otimes \chi_{\epsilon}(w),
\ee
where $\gamma_S = e^{-\beta \HH_S}/Z_{\HH_S}$ is the thermal (Gibbs) state of the system, $Z_{\HH_S}$ is the partition function of the Hamiltonian $\HH_S$, $\beta=1/kT$ is the inverse temperature of the bath and
\be
\label{eq:workbit}
\chi_{\epsilon}(w) = \epsilon \ketbra{0}{0} + (1-\epsilon) \ketbra{1}{1}.
\ee
The arrow in Eq.~\eqref{example} represents a catalytic thermal operation.

It has recently been shown \cite{brandao2013second} that the existence of a catalytic thermodynamic transformation between two states incoherent in energy (as, \emph{e.g.}, in Eq.~\eqref{example}) is \emph{equivalent} to the decrease of a family of generalized free energies, $\{ F_{\alpha}\}$, where $\alpha\geq 0$. These are defined as \mbox{$F_{\alpha}(\xi)= kT S_{\alpha}(\xi\|\gamma) - kT \log Z_{\mathcal{H}}$}, where $\gamma = e^{-\beta \mathcal{H}}/Z_\mathcal{H}$ and $S_{\alpha}(\cdot\|\cdot)$ are informatic-theoretic generalizations of the relative entropy, called R\'enyi divergences (see \cite{renyi1961measures} and Supplemental Material B). $\mathcal{H}$ is the total Hamiltonian of the system (in the case of Eq.~\eqref{example}, \mbox{$\HH=\HH_S + \HH_W$}). For $\alpha \rightarrow 1$, $F_1(\xi)\equiv F(\xi) = \tr{}{\xi \HH} - kT H(\xi)$, the standard non-equilibrium free energy \cite{brandao2011resource, aberg2013truly}, where $H(\xi)$ is the Shannon entropy of $\xi$. Deterministic transformations between non-equilibrium states are severely limited by these constraints. If we look at the asymptotic (or ``thermodynamic") limit in which we process simultaneously a large number $n \rightarrow \infty$ of independent and identically distributed (i.i.d.) states, then all these conditions reduce to the single condition that the free energy $F$ has to decrease~\cite{brandao2011resource}. $F$ also governs processes between equilibrium states, as it coincides with the thermodynamic free energy in those cases: \mbox{$F(\gamma) = -k T\log Z_{\mathcal{H}}$}. However, non-equilibrium thermodynamics of few systems (or many, but correlated) is governed by all $\{F_{\alpha}\}$. Far from equilibrium processes and non-negligible correlations are expected to be common in small-scale thermodynamics, so the $\{F_{\alpha}\}$ are expected be relevant in this regime.

Consider for example the choices $\beta E=1$, $\beta w=0.01$, $p=0.73$, $\epsilon =0.007$ in Eq.~\eqref{example}. The free energy $F$ decreases, $\Delta F < 0$. However for other $F_\alpha$ this is not the case. One can explicitly compute from Eq.~\eqref{example}
\[
\frac{\Delta F_{\alpha}}{kT} =-\log (1+ e^{-\beta E}) + \frac{1}{\alpha -1}\log \frac{\epsilon^{\alpha} + (1-\epsilon)^{\alpha} e^{-\beta w (1-\alpha)}}{p^{\alpha} + (1-p)^{\alpha} e^{-\beta E (1-\alpha)}}
\]
and check that there is a range of $\alpha$'s for which \mbox{$\Delta F_{\alpha}>0$} (\emph{e.g.}, take $\alpha=4$). Hence no catalytic thermal operation can perform the work extraction of Eq.~\eqref{example}, despite $\Delta F <0$.

Reconsider however the transformation in Eq.~\eqref{example}, and now let us use two auxiliary systems $c_1$, $c_2$ with trivial Hamiltonian and ground state occupations equal to $s$ and $q$, respectively. We assume that these get correlated in the process, without changing their local states:
\be
\label{eq:examplecatalysts}
\rho \otimes \ketbra{0}{0} \otimes c_1 \otimes c_2 \longrightarrow \gamma_S \otimes \chi_\epsilon(w) \otimes c_{12}.
\ee
Here $c_{12} $ is the final, correlated state of the two catalysts. Choose $s=0.95$, $q=0.70$, $c_{12}:=(x_{00},x_{01},x_{10},x_{11})= (0.66, 0.29, 0.04, 0.01)$. We assume the process created correlations between the catalysts without changing their marginals, as in Fig.~\ref{fig:correlations}. Hence $c_1 = (x_{00} + x_{01}, 1- x_{00} - x_{01})=(s,1-s)$, and similarly for $c_2=(q,1-q)$. One can check that despite the correlations we still have $\Delta F <0$. Moreover, using the techniques of \cite{horodecki2013fundamental} (i.e.\ looking at the thermomajorization curves for the process of Eq.~\eqref{eq:examplecatalysts}, see Supplemental Material~C), one can prove that there exists a thermodynamic process performing the transformation in Eq.~\eqref{eq:examplecatalysts}. This may seem puzzling: if the process in Eq.~\eqref{example} is impossible, why is \eqref{eq:examplecatalysts} now possible? To understand this, we need to reconsider the notion of entropy for non-equilibrium systems.

\subsection{Anomalous $\alpha$-entropy production}

Non-equilibrium thermodynamics of small systems presents severe challenges, but we can turn some peculiar features of this regime to our advantage. One striking difference between non-equilibrium and equilibrium thermodynamics is that in the latter a unique entropy function exists, characterizing the thermodynamics of the systems at hand \cite{lieb1999physics}. Conversely, the uniqueness of the entropy function is provably equivalent to physical conditions which are very unlikely to be satisfied by non-equilibrium processes \cite{lieb2013entropy}. For example, it implies that given any two arbitrary non-equilibrium states $A$ and $B$ there exists a thermodynamic process connecting $A$ to $B$ or vice versa (``Comparison Hypothesis''). Moreover, it implies a scale-invariance property that can hold only for an effective theory of macroscopic systems \cite{lieb2013entropy}. Therefore the existence of a family of free energies $\{F_{\alpha}\}$ is not a mathematical curiosity with no bearing on physics, but is tightly linked to the fundamental properties of non-equilibrium systems. This gives the multiple constraints of non-equilibrium thermodynamics of \cite{brandao2013second}, but also, as we shall now see, a key, counterintuitive property of correlations: they can generate entropy while being created.

A result of \cite{brandao2013second} is that the catalysts used in the thermodynamic processes can always be chosen to have trivial Hamiltonians. Hence the free energies of the catalysts are given by $F_{\alpha} = -  kT H_{\alpha}$, where $H_{\alpha}$ are information-theoretic generalization of the Shannon entropy called R\'enyi entropies (see \cite{renyi1961measures} and Supplemental Material B). One has $H_1\equiv H$, the Shannon entropy.

Because we usually deal with ``the'' entropy $H$, we have some hard-wired intuitions about the connection between correlations and entropy. For example, we expect two uncorrelated probability distributions to become less disordered when correlations are created (without changing the marginals). Intuitively this is because knowing the realization of one of them allows (due to correlations) to more easily guess the realization of the other. This is captured by the well-known subadditivity of the entropy \cite{shannon1948mathematical} and by the relation 
\be
\label{eq:mutual}
H(p_{AB}) = H(p_A) + H(p_B) - I(p_{AB}),
\ee
where $I(p_{AB})$ (implicitly defined by Eq.~\eqref{eq:mutual}) is the mutual information between $A$ and $B$, and $p_A$, $p_B$ are the marginals of the joint distribution $p_{AB}$. \mbox{$I(p_{AB})\geq 0$} (and \mbox{$I(p_{AB}) = 0$} if and only if \mbox{$p_{AB} = p_A \otimes p_B$}) implies \mbox{$H(p_{AB})<H(p_A \otimes p_B)$} whenever $p_{AB}$ is correlated. It seems that creating correlations has an average work cost~\cite{oppenheim2002approach, reeb2013proving}, because it leads to a reduction of entropy. 

However, as discussed above, for non-equilibrium processes we are forced to use many notions of entropy and some of them are at odds with this intuition. In other words the creation of correlations can be associated to an entropy \emph{production}:
\begin{equation}
H_{\alpha}(p_A\otimes p_B) < H_{\alpha}(p_{AB}).
\label{eqAnomalous}
\end{equation}
We call this property ``anomalous $\alpha$-entropy production''. If, for some $\alpha\neq 1$, Eq.~\eqref{eqAnomalous} holds for some distribution $p_{AB}$,
this suggests that the creation of correlations can ease the thermodynamic transformation. Indeed, we will see that the creation of correlations between the catalysts used in the process massively enlarges the set of accessible states and allows one to extract much more high-quality work than would have been possible otherwise. 

We hinted at the fact that this is due to anomalous $\alpha$-entropy production. The non-uniqueness of entropy carries physical consequences at odds with what is expected in the regimes where one entropy provides a complete description. The following result shows that what we came across in the example of the previous section is a general thermodynamical property.

\subsection{A general result}
\label{general}

Let us denote by $c_1$, \ldots, $c_N$ the marginals of an $N$-partite system $c_{1,\ldots,N}$. The general thermodynamical property is the following: whenever we are given two states that satisfy \mbox{$\Delta F\leq 0$}, we can find auxiliary systems and correlations among them that make the transformation thermodynamically possible:

\begin{thm}
\label{theorem}
Consider a system with Hamiltonian $\HH_S$ and states $\rho$ and $\sigma$ block-diagonal in energy. The three following statements are equivalent:
\begin{enumerate}
\item There exists a thermodynamic process transforming $\rho$ into a state $\sigma_{\epsilon}$ arbitrarily close to $\sigma$, by creating correlations among auxiliary systems, but without changing their local states:
\be
\label{eq:activation}
\rho \otimes c_{1} \otimes \cdots \otimes c_N \rightarrow \sigma_{\epsilon} \otimes c_{1,\ldots,N}.
\ee
One can always choose $N \leq 3$ and trivial Hamiltonians for the auxiliary systems.
\item There exists $c_1$,\ldots,$c_N$ and $c_{1,\ldots,N}$ such that anomalous $\alpha$-entropy production ensures that all $\{F_{\alpha}\}$ constraints are satisfied
in Eq.~\eqref{eq:activation}.
\item $F(\rho) \geq F(\sigma)$.
\end{enumerate}
\end{thm}

For a rigorous statement and proof, see Supplemental Material D. The proof is based on a generalization of the notion of catalytic majorization introduced in \cite{mueller2015non}. Theorem~\ref{theorem} says that whenever a transformation is possible in the thermodynamic limit (i.e.\ when \mbox{$\Delta F\leq0$}, see \cite{brandao2011resource}), then it is also possible by processing individual systems in the single-shot regime; what is needed is the creation of correlations among auxiliary systems whose local state is left unchanged. This is a surprising simplification of the thermodynamic ordering, compared to the infinite constraints $\Delta F_{\alpha} \leq 0$ of~\cite{brandao2013second}, and provides a non-asymptotic, operational meaning to the non-equilibrium free energy $F$.  

It is useful to compare with recent results on work extraction from single quantum systems. The free energy $F$ gives absolute limits on the average amount of energy that can be extracted from single systems out of equilibrium \cite{skrzypczyk2014work}. However for small, single systems, the work distribution can be very broad. These fluctuations are a function of the initial non-equilibrium state and can be of the same order as the average extracted energy itself. Hence, arguably, the energy extracted can be more heat-like than work-like \cite{aberg2013truly}. 

Since the ability to extract fluctuation-free work seems crucial for any engine that is trying to operate reliably in a non-equilibrium environment, deterministic work extraction has been recently investigated in \cite{horodecki2013fundamental, brandao2013second}. It has been shown that from a system $\rho$ incoherent in energy we can deterministically extract work equal to $F_0(\rho)<F(\rho)$. However, it holds $F_0(\rho) = 0$ for any full-rank state. As we can never ensure experimentally that a state does not have full rank, this immediately implies that strictly deterministic work extraction through thermal operations is practically impossible; at best, the work yield will become tiny compared to $F(\rho)$ the smaller the failure probability we tolerate. This is why the authors of \cite{horodecki2013fundamental} allow for some fixed (as opposed to arbitrarily small) error probability in their model, as does, similarly, \AA berg in his analysis~\cite{aberg2013truly}. Indeed, as \AA berg's model shows, the role of the error probability is to focus on sufficiently likely energy levels of the system, a safeguard from unlikely but potentially harmful energy fluctuations.

The considerations above suggest that when performing work extraction at the nanoscale we either extract very
little or no work, or we must include some large enough error probability in the protocol. The twist of the present result is that neither of the two is necessary. Error-free, fluctuation-free work extraction from non-equilibrium systems is possible at optimal output
$F(\rho)-F(\gamma_S)$ by creating correlations in the auxiliary systems used in the process. From Theorem \ref{theorem} it immediately follows that given $\rho$ one can
find auxiliary systems in a state $c_{1,\ldots,N}$, $N \leq 3$, such that
\be
\label{eq:workextraction}
\rho \otimes \ketbra{0}{0} \otimes c_1 \otimes \cdots \otimes c_N \rightarrow \gamma_S \otimes \chi_{\epsilon}(w) \otimes c_{1,\ldots,N},
\ee
where $w= F(\rho)-F(\gamma_S)$, and $\epsilon>0$ can be chosen to be \emph{arbitrarily} close to zero. Physically speaking, there is no difference between protocols ensuring an arbitrarily small error probability in the work bit $\chi_{\epsilon}(w)$ and a deterministic protocol that gives the pure excited state $\ket{1}$ (see Supplemental Material E for a comparison with other work extraction models). Also notice that we only need to build up an arbitrarily small amount of correlations among the auxiliary systems, as measured by the mutual information. Indeed one can easily check from the non-increase of $F$ in~(\ref{eq:workextraction}) that
$I(c_{1,\ldots,N})\leq H(\epsilon,1-\epsilon)+\epsilon\, w/(kT)$,
where $I$ generalizes mutual information to $N$-partite systems (also known as total correlation \cite{watanabe1960information, vedral2002role}):
\be
\nonumber
I(c_{1,\ldots,N}) = S_1\left(c_{1,\ldots,N}\left\|\bigotimes_{i=1}^N c_i\right.\right) = \sum_{i=1}^N H(c_i) - H(c_{1,\ldots,N}).
\ee
Actually, we also show that arbitrarily small $I(c_{1,\ldots,N})$ can be achieved for any transformation in this framework, not only for work extraction
(see Supplemental Material F). This is why Theorem~\ref{theorem} gets around the limitations analyzed in~\cite{oppenheim2002approach,reeb2013proving}.

\subsection{Conclusions}

The non-equilibrium free energy $F$ is known to have meaning for the thermodynamics of a large number of uncorrelated systems~\cite{brandao2011resource}. This is not surprising, because $F(\rho)= \tr{}{\rho \HH} - kT H(\rho)$, and the von Neumann entropy $H$ acquires its operational meaning in tasks involving infinitely many, identical states~\cite{nielsen2010quantum, dahlsten2011inadequacy}. However we have shown here that $F$ has a novel operational meaning for single-shot thermodynamics, i.e.\ for non-equilibrium, irreversible transformations on single systems.
 
An engine operating on correlated systems out of equilibrium will face extra irreversibility in comparison to the asymptotic or equilibrium regimes. For example, the amount of work $w_{\rm{form}}$ necessary to form a state exceeds the amount of work $w_{\rm{ext}}$ that can be extracted from it. This is because $w_{\rm{form}} = F_{\infty} > F_{0} = w_{\rm{ext}}$ \cite{horodecki2013fundamental, brandao2013second}. However Theorem \ref{theorem} shows that an engine can operate at the reversible limit $w_{\rm{form}}=w_{\rm{ext}}=F$ if it can build up a small amount of correlations among few auxiliary systems. In principle, we can think of an engine operating on out-of-equilibrium microscopic systems and accessing a resource of ``stochastic independence'' (i.e.\ uncorrelated, auxiliary systems). 

An interesting open question is the following: can we engineer typical (weak) system-environment interactions to generate the required correlations among the auxiliary systems? This would allow for a practical implementation of the engine. Moreover, are $N=2$ auxiliary systems always enough? Can we give any bound on their sizes, as for standard catalysis~\cite{ng2014limits}? One can show, for example, that a transition where $F$ is constant can be performed creating an arbitrarily small amount of correlations only if the dimension of the auxiliary systems grows without bound (see Supplemental Material G).

It is straightforward to generalize this result to the situation in which the Hamiltonian of the initial state and that of the target state do not coincide, using the results of \cite{horodecki2013fundamental, brandao2013second}. A more difficult question is to ask what happens in the case in which the states are not block-diagonal in energy. Then considerations involving free energies do not suffice \cite{lostaglio2015description,cwiklinski2014towards, lostaglio2015quantum} and asymmetry measures $\{A_{\alpha}\}$ quantifying the coherent content of the systems pose further constraints. Does the creation of correlations help also in this case? One could hope so, as the creation of correlations can ``hide'' local coherence, \emph{e.g.} coherence can be created locally with energy-preserving transformations if correlations between systems are created \cite{aberg2014catalytic}. We leave this open for future research.

It also seems worthwhile to look for concrete physical situations where local states $c_i$ of large quantum systems, interacting with other systems in a heat bath, are forced to remain constant (say,
due to local conservation laws). Our result suggests that there could be a tendency to build up correlations,
similarly as there is a tendency to thermalize if the purity of the local states is allowed to decrease.
This is particularly interesting due to the fact that the transition from product to correlated states is often
regarded as an instance of an arrow of time.

\bigskip

\textbf{Acknowledgments.} ML would like to thank D.\ Jennings, T.\ Rudolph, K.\ Korzekwa and A.\ M.\ Alhambra for helpful comments on a draft of this work, and M.\ P.\ Woods for a useful comment concerning the dimension of the auxiliary systems. ML is supported in part by EPSRC and COST Action MP1209. MM and MP are grateful to F.\ Brand\~ao, G.\ Gour, and V.\ Narasimhachar for comments on an earlier draft. Research at Perimeter Institute is supported by the Government of Canada through Industry Canada and by the Province of Ontario through the Ministry of Research and Innovation. MP thanks the Heidelberg Graduate School of Fundamental Physics for financial support.

\bibliography{Bibliography_thermodynamics_4}

\onecolumngrid

\section{Appendix}
\subsection{Thermal operations and catalysis}
\label{operations}

This work is based on the resource-theory approach to quantum thermodynamics, introduced in \cite{janzing2000thermodynamic, brandao2011resource} and inspired by developments in the theory of quantum entanglement \cite{horodecki2009quantum}. The basic idea is to define a set of restrictions on the allowed operations on a system. This in turn distinguishes between states that can be freely prepared given the restrictions from those that are not and are hence resourceful. In thermodynamics, given a system $S$ with Hamiltonian $\mathcal{H}_S$ in state $\rho$, the set of allowed transformations is given by all energy-preserving unitary interactions $U$ between the system and a thermal bath with an arbitrary Hamiltonian $\mathcal{H}_B$ and fixed temperature $T$, and subsequent partial trace:
\be
\label{eq:thermalmap}
\mathcal{E}(\rho) = \tr{B}{U(\rho \otimes \gamma_B)U^{\dag}}, \quad [U, \HH_S + \HH_B]=0,
\ee
where $\gamma_B = e^{-\beta \HH_B}/{Z_B}$, $Z_B = \tr{}{e^{-\beta \HH_B}}$, $\beta=1/kT$. These transformations have been called \emph{thermal operations}. The only state that we can prepare for free without trivializing the theory is the thermal Gibbs state of the system \cite{brandao2013second}, $\gamma_S = e^{-\beta \HH_S}/{Z_S}$, $Z_S = \tr{}{e^{-\beta \HH_S}}$. This set was extended in \cite{brandao2013second} to allow for the use of catalysts, i.e.\ auxiliary states $c$ that activate an otherwise forbidden transformation, but are given back unchanged and uncorrelated. We say that there exists a \emph{catalytic thermal operation} mapping $\rho$ to $\sigma$, and write $\rho \rightarrow \sigma$, if there is a catalyst $c$ and a thermal operation mapping $\rho \otimes c$ to $\sigma \otimes c$. This framework can be extended to the case where the Hamiltonian is allowed to change,
by introducing a clock degree of freedom~\cite{horodecki2013fundamental, brandao2013second}. 

Thermodynamics is tightly linked, at least in the case of states with no quantum coherence, to the resource theory of nonuniformity \cite{gour2013resource}, through an embedding that resembles the connection between microcanonical and canonical ensembles in statistical mechanics \cite{brandao2013second, lostaglio2015description}. This point will be crucial in proving our main result.

\subsection{$\alpha$-free energies and R\'enyi entropies}
\label{renyidefinition}

Let $\rho$ be the state of a system with Hamiltonian $\mathcal{H}$ and $\gamma$ the correspondent thermal state. Then the \mbox{$\alpha$-free} energy of $\rho$ is defined as
\[
F_{\alpha}(\rho) = - kT \log Z_{\mathcal{H}} + k T S_{\alpha}(\rho\|\gamma),
\]
where $S_{\alpha}$ are the so-called $\alpha$-R{\'e}nyi divergences \cite{wilde2013strong, muller2013quantum}. Notice that if $\rho = \gamma$ is an equilibrium state, then $F_{\alpha}$ coincides with the thermodynamic free energy. Because in this work we focus on quantum states block-diagonal in the energy eigenbasis, i.e.\ $[\rho, \gamma]=0$, the definition of $S_{\alpha}$ reduces that the one given by R{\'e}nyi himself. Denoting the eigenvalues of $\rho$ and $\gamma$ by $p_i$ resp.\ $q_i$, we get
$S_{\alpha}(\rho\|\gamma)= S_{\alpha}(p\|q)$, and the latter is defined for $\alpha\in\R\setminus\{0,1\}$ as~\cite{renyi1961measures}
\[
S_{\alpha}(p\|q) =  \frac{{\rm sgn}(\alpha)}{\alpha-1}\log \sum_i p^{\alpha}_i q^{1-\alpha}_i.
\]
The cases $\alpha\in\{-\infty,0,1,+\infty\}$ are defined via suitable limits (see e.g.~\cite{brandao2013second}):
\be
\nonumber
S_{\infty}(p\|q) = \log \max_i p_i/q_i, \quad S_1(p\|q) = \sum_i p_i \log (p_i/q_i), 
\ee
\be
\nonumber
S_0(p\|q) = - \log \sum_{i | p_i \neq 0} q_i, \quad S_{-\infty}(p\|q) = S_\infty(q\|p).
\ee
R{\'e}nyi also defined $\alpha$-entropies as follows:
\[
H_{\alpha}(\rho) = \frac{{\rm sgn}(\alpha)}{1-\alpha}\log \sum_i p^{\alpha}_i.
\]
These are linked to the notion of trumping (that is, catalytic majorization) that found application in the theory of entanglement \cite{klimesh2007inequalities, turgut2007catalytic}. The cases $\alpha \in \{-\infty,0,1,+\infty \}$ are also defined by appropriate limits. Denoting by $H$ the Shannon entropy,
\begin{eqnarray*}
H_\infty(p) &=& - \log \max_i p_i, \quad H_1(p) = -\sum_i p_i \log p_i \equiv H(p),\\
H_0(p) &=& \log {\rm rank}(p), \quad H_{-\infty}(p) = \log \min_i p_i.
\end{eqnarray*}

Using the expression above for $S_1$, one can check the $\alpha=1$ free energy can be rewritten as
\[
F_1(p) = \langle \HH \rangle - kT H(p) \equiv F(p).
\]
This free energy is known to characterize transformations between a large number of identical and uncorrelated systems \cite{brandao2011resource}.

All the free energies $\{F_{\alpha}\}$ are monotonically decreasing under thermal operations. In fact, thermal operations have the Gibbs state as a fixed point, \mbox{$\mathcal{E}(\gamma)=\gamma$}, and the relative entropies $S_{\alpha}$ satisfy the data-processing inequality: for every $p$, $q$ and stochastic map $\Lambda$, \mbox{$S_{\alpha}(\Lambda(p) \| \Lambda (q))\leq S_{\alpha}(p \| q) $} \cite{erven2007renyi}. From this one can easily see that for any block-diagonal state $\rho$,
\be
\label{eq:monotonicity}
\mathcal{E} \textrm{   thermal op.   } \Rightarrow F_{\alpha}(\mathcal{E}(\rho)) \leq F_{\alpha}(\rho), \quad \forall \alpha \in \R.
\ee
This result can be generalized to arbitrary quantum states, using extensions of the definition of $F_{\alpha}$ and the data-processing inequality \cite{brandao2013second} to the fully quantum case. The main result of~\cite{brandao2013second} is to prove that the condition 
\be
\label{eq:entropicconditions}
F_{\alpha}(\rho) \geq F_{\alpha}(\sigma) \quad \mbox{for all } \alpha \in \R
\ee
is actually \emph{sufficient} for the existence of a catalytic thermal transformation between $\rho$ and (an arbitrarily good approximation of) $\sigma$ when either of them is block-diagonal in energy. However, (\ref{eq:entropicconditions}) is not sufficient to characterize all possible transformations between states with coherences between energy levels~\cite{lostaglio2015description}.

It is possible to check directly that for the $d$-dimensional maximally mixed state $\eta$, it holds
\[
S_{\alpha}(p\| \eta) = \log d - H_{\alpha}(p).
\]
This implies that for trivial Hamiltonians $\mathcal{H}\equiv 0$,
\be
\label{eq:trivial}
F_{\alpha}(p) = -kT H_{\alpha}(p).
\ee

\subsection{Numerical example}

We present here the numerical example of the main text, alongside with the theory necessary to understand its implications. As in the rest of the paper, the notation ``$\rightarrow$'' denotes a catalytic thermal operation (see Appendix A).

Our goal is to find two states $\rho$ and $\sigma$
such that $\rho\rightarrow\sigma$ is forbidden by the laws of thermodynamics, but such that the transition becomes possible
when we allow the presence of correlations in the final state of the catalysts. We will look for states that violate the condition $F_{\alpha}\left(\rho\right)\geq F_{\alpha}\left(\sigma\right)$ for some $\alpha \neq 1$, but such that the free energy $F\equiv F_1$ decreases in the process. The former condition ensures that $\rho \rightarrow \sigma$ is impossible (see Appendix B and \cite{brandao2013second}). The latter condition is required, because even if we allow correlations to be created among the catalysts $c_1$,\ldots,$c_N$,
the condition $\Delta F:=F(\sigma)-F(\rho)\leq 0$ remains necessary for the transformation to be possible.
Specifically, if $c_{1,\ldots,N}$ is a correlated state with marginals equal to $c_1$,\ldots,$c_N$, one gets
\[
\rho \otimes c_1 \otimes \cdots \otimes c_N \rightarrow \sigma \otimes c_{1,\ldots,N} \Rightarrow F(\rho) \geq F(\sigma),
\]
where we used the monotonicity and superadditivity of the free energy $F$.  

Since we are dealing with states block-diagonal in the energy eigenbasis, we can substitute the density matrices
of every state with the probability distributions over energy. Hence, we only need to find a pair of probability distributions
$p$ and $q$ such that the transition $p\rightarrow q$ is
not allowed, but there are auxiliary systems $c_{1},\ldots,c_{N}$
such that $p\otimes c_{1}\otimes c_{2}\otimes\cdots\otimes c_{N}\rightarrow q\otimes c_{1,\ldots,N}$
becomes possible. Indeed, we will only need $N=2$ for our example.

A simple way to check whether there is a thermal operation
connecting two states $\xi_1$ and $\xi_2$ block-diagonal in the
energy eigenbasis is to verify that $\xi_1$ thermomajorizes $\xi_2$
\cite{brandao2013second}. Thermomajorization can be viewed as an extension of the notion of majorization \cite{ruch1978mixing, horodecki2013fundamental, gour2013resource}, a quasi-order between probability distributions that finds many applications, from economics to the theory of quantum entanglement \cite{horodecki2009quantum}. In the next subsection, we introduce the notion of thermomajorization, and in the last subsection we use it to find an example of a creation of correlations that allows an otherwise impossible thermodynamic transformation.

\subsubsection*{From majorization to thermomajorization}

For the purpose of this discussion we introduce majorization geometrically, through the notion of Lorenz curve (an equivalent, algebraic, definition is given in Appendix D). Let us consider a system in a state $\xi_1$. Denote by $p$ the probability distribution given by the eigenvalues of the density matrix $\xi_1$. From this
probability distribution we can define the vector $p^{\downarrow}$
whose elements are given by the entries of $p$ in decreasing order
($p_{1}^{\downarrow}\geq p_{2}^{\downarrow}\geq\ldots\geq p_{m}^{\downarrow}$). Now we can build a cartesian plot of a piecewise linear function whose $n$-th point is given
by $\left\{n,\sum_{i=1}^{n}p_{i}^{\downarrow}\right\}$; this plot is called
the Lorenz curve of $\xi_1$ \cite{marshall2010inequalities}. We can do the same for the
probability vector $q$ of the eigenvalues
of another state $\xi_2$ that has the same dimension as $\xi_1$.
Once we have these two piecewise linear plots, we say that $\xi_1$ majorizes $\xi_2$ (and write $\xi_1 \succ \xi_2$) when the Lorenz curve of $\xi_1$ is everywhere on or above the Lorenz curve of $\xi_2$. $\xi_1 \succ \xi_2$ can be proved to be equivalent to the existence of a so-called noisy operation achieving the transition from $\xi_1$ to $\xi_2$ \cite{horodecki2003reversible}. The connection with thermodynamics comes from the fact that noisy operations can be thought of as thermal operations (introduced in Appendix A) with trivial Hamiltonians~\cite{gour2013resource}.

The majorization criterion can be extended to the case of systems
with nontrivial Hamiltonians and finite temperatures \cite{ruch1978mixing, horodecki2013fundamental}. Let us consider a system in a state $\xi_1$ block-diagonal in energy, with eigenvalues $p_{i}$ representing the probability that the system is in the $i$-th state of energy $E_{i}$. We want to transform this state into a target state $\xi_2$ by coupling the system to a thermal bath at inverse temperature $\beta = 1/kT$, i.e.\ we ask if there is a thermal operation transforming $\xi_1$ to $\xi_2$. 

To check if this is possible, the simple ordering of majorization is now replaced by the notion of $\beta$-ordering \cite{horodecki2013fundamental}. Define by $g_i$ the vector of Gibbs factors, $g_i=e^{-\beta E_{i}}$. We construct the vector whose elements are Gibbs-rescaled probabilities, $p_{i}/g_i$, and sort its elements in decreasing
order. If for two indices $i$ and $j$, the rescaled probabilities coincide, $i$ will precede $j$ if $p_i \geq p_j$. Using this ordering, we obtain the so-called $\beta$-ordered version of $p$, denoted by $p^{\downarrow \beta}$. With the same ordering, we obtain the vector $g^{\downarrow \beta}$. We can now construct the thermal Lorenz curve of $\xi_1$, that is
the plot of a piecewise linear function whose $n$-th point is $\left\{\sum_{i=1}^{n}g^{\downarrow \beta}_{i},\sum_{i=1}^{n}p^{\downarrow \beta}_{i}\right\}$. We can follow the same procedure for $\xi_2$. We say that $\xi_1$ thermomajorizes $\xi_2$ if the thermal Lorenz curve of $\xi_1$ is everywhere on or above the thermal Lorenz curve of $\xi_2$. Notice that, for $\beta =0$ (infinite temperature) or $E_i=0$ for all $i$ (trivial Hamiltonian), $p^{\downarrow \beta} = p^{\downarrow}$ and $g^{\downarrow \beta}_{i}\equiv 1$, so we recover majorization. 

\begin{figure}[t!]
\includegraphics[width=0.5\columnwidth]{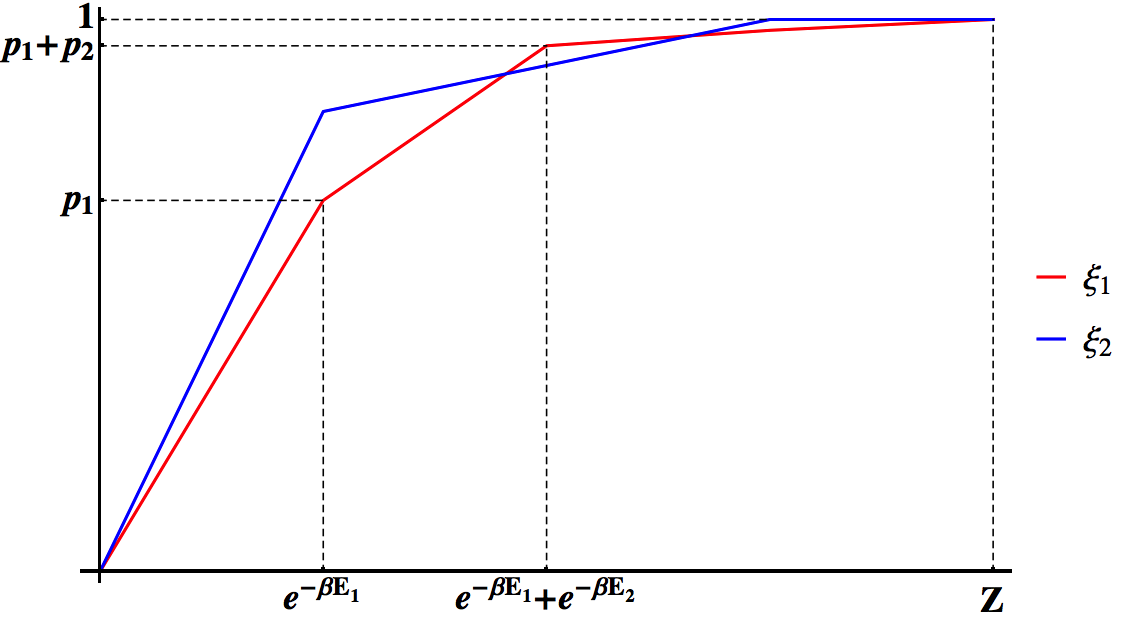}
\caption{Thermo-majorization: the red and blue thermal Lorenz curves represent two states. There is no 
thermal operation converting one into the other, because the curves are crossing.}
\label{fig:thermo_majorization}
\end{figure}

The crucial result is that a thermal operation from $\xi_1$ to $\xi_2$ (block-diagonal in energy) is possible if
and only if the thermal Lorenz curve of $\xi_1$ is everywhere on or above
the thermal Lorenz curve of $\xi_2$
\cite{horodecki2013fundamental}. In some cases, it happens that the thermal Lorenz curves are not disjoint (see Fig.~\ref{fig:thermo_majorization}).
This means that there is no thermal operation that can
achieve the transition from $\xi_1$ to $\xi_2$ (nor from $\xi_2$ to $\xi_1$). In this case we can try
to find an auxiliary system $c$ such that the thermal Lorenz curve of $\xi_1 \otimes c$ lies everywhere on or above
the thermal Lorenz curve of $\xi_2 \otimes c$.
This means to look for a catalyst for our transition, a system that allows an otherwise impossible transition and yet is left
unchanged at the end of the process. As seen in Appendix B, there exist necessary and sufficient conditions for this to be possible (Eq.~\eqref{eq:entropicconditions}).

\subsubsection*{Finding a correlating catalytic operation}

With the thermomajorization criterion in mind, we reconsider the task introduced at the beginning of this Appendix: to find
a pair of states $\xi_1$ and $\xi_2$ such that $\xi_1 \rightarrow \xi_2$ is impossible, but becomes possible if we allow to build up
correlations among the catalysts (for brevity, we will call this a \emph{correlating-catalytic operation} from $\xi_1$ to $\xi_2$). We can now reformulate the problem of finding a non-trivial correlating catalytic operation as follows: find $\xi_1$ and $\xi_2$ s.t. $F_{\alpha}(\xi_1) < F_{\alpha}(\xi_2)$ for some $\alpha \neq 1$, but such that $F(\xi_1)>F(\xi_2)$; then find uncorrelated systems $c_1,\ldots,c_N$ and some correlations between them (that is, a state $c_{1,\ldots,N}$ with marginals $c_1$,\ldots,$c_N$), such that the thermal Lorenz curve of $\xi_1 \otimes c_1 \otimes \cdots \otimes c_N$ lies all above the thermal Lorenz curve of $\xi_2 \otimes c_{1,..,N}$. Since all involved states are block-diagonal in energy, we identify in the notation quantum states and the vectors of their eigenvalues.

With this in mind, consider a system with Hamiltonian \mbox{$\mathcal{H}_{S}=E\ketbra{1}{1}$}, in an initial state $\rho$ block-diagonal in the energy basis. The occupation probabilities of the ground and excited state are denoted by \mbox{$\left(p,1-p\right)$}. We consider the work extraction process introduced in the main text:
\be
\label{eq:impossibleex}
\rho \otimes \ketbra{0}{0} \longrightarrow \gamma_S \otimes \chi_\epsilon(w),
\ee
where a work bit with Hamiltonian \mbox{$\mathcal{H}_{W}=w\ketbra{1}{1}$} and initially in the ground state gets $\epsilon$-close to the excited state. Notice that in Eq.~(\ref{eq:impossibleex}) we used the same notation of the main text, where
\[
\chi_{\epsilon}(w) = \epsilon \ketbra{0}{0} + (1-\epsilon) \ketbra{1}{1}.
\]

In order to maximize work extraction, we assume that the system is, at the end of the process, in a thermal state
$\gamma_{S}=e^{-\beta \mathcal{H}_S}/ Z_S$, with $Z_S=1+e^{-\beta E}$ the partition function of $\mathcal{H}_S$. 
This is summarized in Table~I.

\begin{table}[b!]
\label{table}
\begin{centering}
\begin{tabular}{c|c|c|c|c|c|}
\cline{2-6}
 & System & Work bit & Cat. (overall) & Cat. 1 & Cat. 2 \tabularnewline
\hline 
\hline 
\multicolumn{1}{ |c|  }{Hamiltonian} & $E\ketbra{1}{1}$ & $w\ketbra{1}{1}$ & $\mathbb{I} \otimes \mathbb{I}$ & $\mathbb{I}$ & $\mathbb{I}$\tabularnewline
\hline 
\multicolumn{1}{ |c|  }{Initial state} & $\rho$ & $\ketbra{0}{0}$ & $c_{1} \otimes c_2$ & $c_{1}$ & $c_{2}$\tabularnewline
\hline 
\multicolumn{1}{ |c|  }{Final state} & $\gamma_{S}$ & $\chi_{\epsilon}(w)$ & $c_{12}$ &$c_1$ & $c_{2}$\tabularnewline
\hline 
\end{tabular}
\par\end{centering}
\caption{Summary of the states of each system and their Hamiltonians. The two catalysts
$c_{1}$ and $c_{2}$ have trivial Hamiltonians and they are added
with the purpose of making the transition possible. They get correlated, but their local states do not change and they are uncorrelated from system and work-bit.}
\end{table}

\begin{figure}[t!]
\includegraphics[width=0.4\columnwidth]{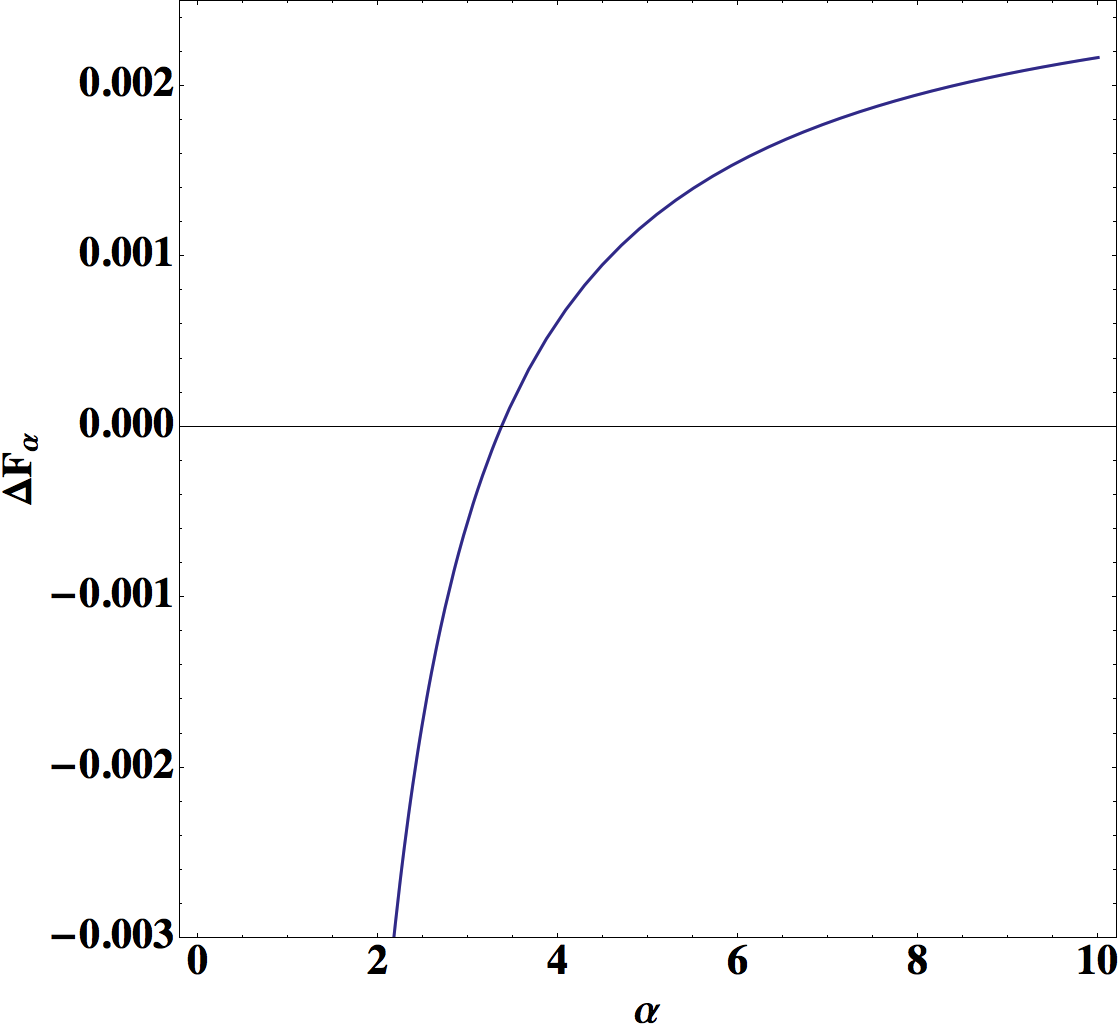}
\caption{The plot of $\Delta F_{\alpha}$ shows explicitly that there exists no catalytic thermal operation allowing the transition from $\rho\otimes\ketbra{0}{0}$ to $\gamma_{S}\otimes\chi_{\epsilon}$. The free energy $F_{1}$ decreases in the process, but this is not the case for all $F_{\alpha}$.}
\label{fig:diff_free_en}
\end{figure}

We now make the choices
\mbox{$\beta E=1$}, \mbox{$\beta w=0.01$}, \mbox{$p=0.73$}, \mbox{$\epsilon=0.007$}. It is then clear from Fig.~\ref{fig:diff_free_en}
that \mbox{$\Delta F_{\alpha}=F_{\alpha}\left(\gamma_{S}\otimes\chi_{\epsilon}\right) - F_{\alpha}\left(\rho\otimes\ketbra{0}{0}\right)$}
is negative for \mbox{$\alpha=1$}, but this not the case for all \mbox{$\alpha>1$}.
According to the results of \cite{brandao2013second}, this means that the transition of Eq.~\eqref{eq:impossibleex} is forbidden,
i.e.\ impossible to achieve by a catalytic thermal operation,
for the chosen set of parameters. Hence, this example is a possible candidate for the construction of a correlating-catalytic transformation.  

\begin{figure}[b!]
\includegraphics[width=0.5\columnwidth]{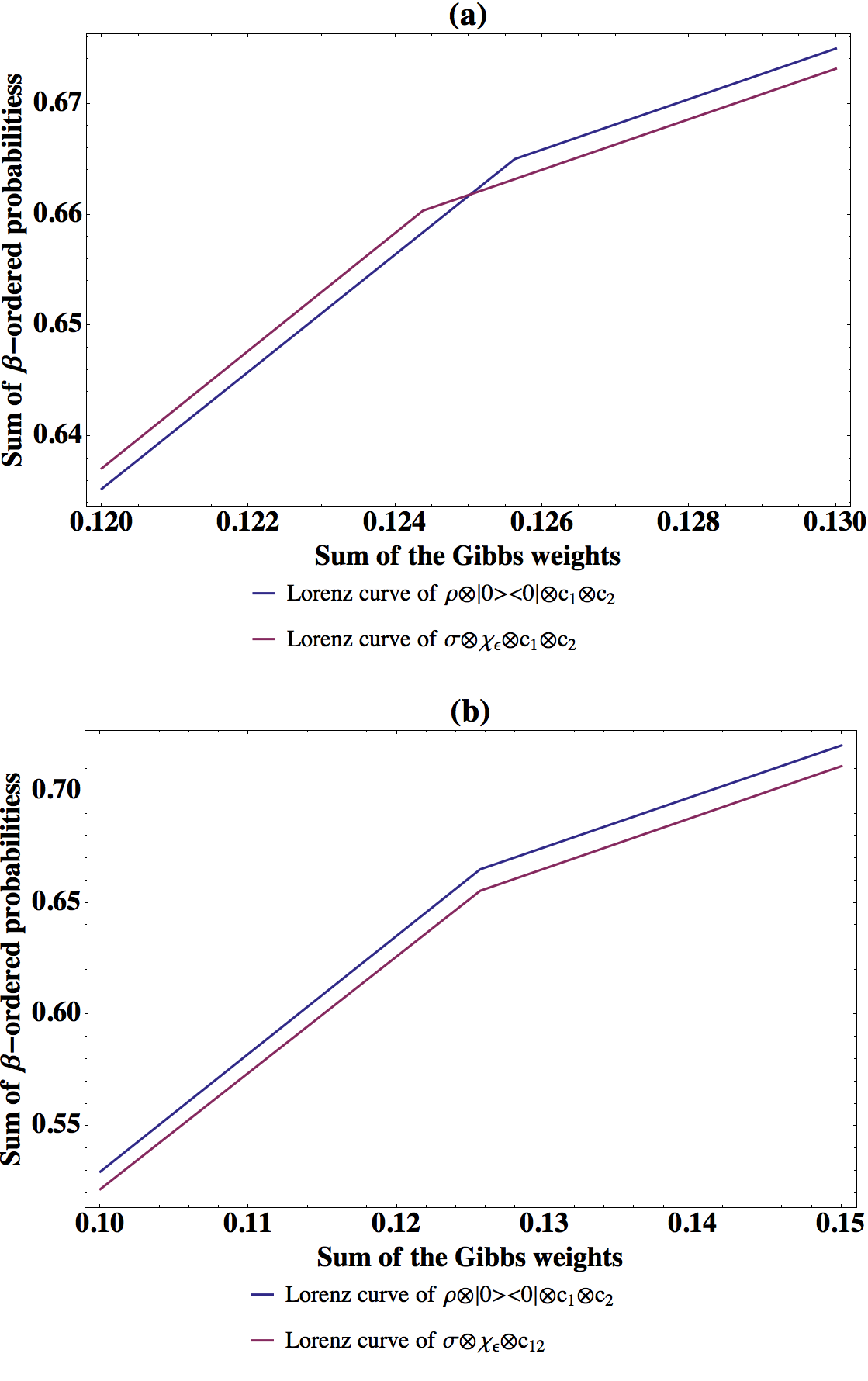}
\caption{Thermal Lorenz curves for the transition in Eq.~\eqref{eq:transition}. In (a) we use two catalysts $c_{1}$ and $c_{2}$ as defined in the text, but they are left uncorrelated in the final state: in this case, the Lorenz curve of the initial and of the final state cross. In (b) we allow the two catalysts to be correlated in the final state: in this case the two Lorenz curves are disjoint and the transition is possible.}
\label{fig:lorenz_curve}
\end{figure}

We consider two qubit auxiliary systems with trivial Hamiltonians.
The vector of their eigenvalues is denoted by $c_{1}=\left(s,1-s\right)$ and $c_{2}=\left(q,1-q\right)$. We choose $s=0.95$ and $q=0.7$. In order to have a correlating-catalytic
transition, we need to check whether there exists correlations such that the following transformation is possible:
\be
\label{eq:transition}
\rho \otimes\ketbra{0}{0}\otimes c_{1}\otimes c_{2} \rightarrow \gamma_{S}\otimes\chi_{\varepsilon}\otimes c_{12}.
\ee
In particular it will suffice to check that \mbox{$\rho \otimes\ketbra{0}{0}\otimes c_{1}\otimes c_{2}$} thermomajorizes \mbox{$\gamma_{S}\otimes\chi_{\varepsilon}\otimes c_{12}$} for an appropriate choice of the correlations in $c_{12}$. Denote by \mbox{$c_{12}:=\left(x_{00},x_{01},x_{10},x_{11}\right)$} the joint probability
distribution of the final state of the two catalysts. We want to impose that its marginals are $c_{1}$ and
$c_{2}$; this means that the catalysts are left unchanged locally (the transition is summarized in Table 1). This condition fixes all but one of the parameters of $c_{12}$, so that we have a one-parameter family of possible correlations, defined as the solutions of the following equations: 
\begin{eqnarray*}
x_{ij} \geq 0, \quad x_{00}+x_{01}+x_{10}+x_{11}=1, \\
x_{00}+x_{01}=s,  \quad x_{00}+x_{10}=q.
\end{eqnarray*}
We solve for $x_{10}$ and choose $x_{10} = 0.04$. Hence we get
\begin{eqnarray*}
x_{00} & = & q-x_{10}=0.66,\\
x_{01} & = & x_{10}+s-q=0.29,\\
x_{11} & = & 1-s-x_{10}=0.01.
\end{eqnarray*}
We now construct the thermomajorization curves for this transition. The result is presented in Fig.~\ref{fig:lorenz_curve}. The initial
state thermomajorizes the final state, so it is possible to
find a thermal map achieving a transition that was provably impossible without building up the correlations among the catalysts.

\subsection{Proof of the main theorem}

The structure of the proof is as follows. First we recall a recent result on the theory of majorization~\cite{mueller2015non} and show that it immediately implies the general result of Theorem~1 in the special case in which the Hamiltonians of all systems involved are trivial. We use then techniques introduced in \cite{brandao2013second} to extend the result to thermodynamics with general Hamiltonians. In this work we assume that all systems are finite-dimensional.

\subsubsection{Result for trivial Hamiltonians}
\label{trivialhamiltonians}

The first step is to recall a recent characterization of von Neumann entropy through a generalized notion of majorization introduced in~\cite{mueller2015non}. We begin with a notion of majorization that can be easily seen to be equivalent to the one above based on Lorenz curves:

\begin{defn}[Majorization~\cite{marshall2010inequalities}]
For classical probability distributions $p=(p_1,\ldots,p_m)$ and $q=(q_1,\ldots, q_m)$, we say that $p$ majorizes $q$, and write $p\succ q$,
if and only if
\[
\sum_{i=1}^k p_i^{\downarrow}\geq \sum_{i=1}^k q_i^{\downarrow}
\]
for all $k=1,2,\ldots,m$, where $p_1^{\downarrow}\geq p_2^{\downarrow}\geq\ldots \geq p_m^{\downarrow}$
denotes the components of $p$ arranged in non-increasing order. For quantum states $\rho$ and $\sigma$, we write $\rho\succ\sigma$ if and only if $\lambda(\rho)\succ\lambda(\sigma)$,
where $\lambda(\rho)$ and $\lambda(\sigma)$ are the probability distributions of eigenvalues of $\rho$ and $\sigma$.
\end{defn}

We now recall some results from~\cite{mueller2015non}. While the results there have originally been formulated for classical probability distributions,
it is easy to see that they carry over directly to quantum states, by identifying the vector of eigenvalues with a probability distribution.

\begin{defn}[c-trumping, \cite{mueller2015non}]
We say that that \emph{$\rho$ c-trumps $\sigma$}, and write $\rho\succ_c\sigma$, if and only if there exists $N\in\mathbb{N}_0$ and a $N$-partite
quantum state $c_{1,2,\ldots,N}$, with marginals $c_1$,\ldots,$c_N$, such that
\be
   \rho\otimes \left(\strut c_1\otimes \cdots \otimes c_N\right)\succ \sigma\otimes c_{1,2,\ldots,N}.
   \label{eqCTrumping}
\ee
\end{defn}
Notice that majorization is a special case of c-trumping when $N=0$ and trumping \cite{klimesh2007inequalities, turgut2007catalytic} is a special case when $N=1$.
We can now report one main result of~\cite{mueller2015non}:

\begin{thm}
\label{Thm1Main}
Suppose that $\rho$ and $\sigma$ do not have identical sets of eigenvalues.
Then $\rho\succ_c\sigma$ if and only if \mbox{${\rm rank}(\rho)\leq{\rm rank}(\sigma)$} and $H(\rho)<H(\sigma)$, for $H$ the von Neumann entropy. Moreover, we can always choose $N=3$ in~(\ref{eqCTrumping}).
\end{thm}

We can now show that the result of \cite{mueller2015non} immediately implies a result valid for thermodynamics, when the Hamiltonians of all systems are trivial. We denote by $\Vert \cdot \Vert$ the trace norm $\Vert X \Vert = \tr{}{\sqrt{X^{\dag}X}}$. However note that the following result actually holds for any general norm, as all norms are equivalent in finite dimension.

\begin{thm}
\label{th:trivialh}
Consider a system with trivial Hamiltonian. The following statements are equivalent:

\begin{enumerate}
\item  For every $\epsilon >0$ there exists a thermal operation $\mathcal{E}_{\epsilon}$ and $N$ auxiliary systems with trivial Hamiltonians and joint state $c_{1,\ldots,N}$
with marginals $c_1,\ldots,c_N$ such that
\begin{enumerate}
\item  $\Vert \sigma - \sigma_{\epsilon}\Vert<\epsilon$,
\item  $\mathcal{E}_{\epsilon}(\rho \otimes c_{1} \otimes \cdots \otimes c_N)= \sigma_{\epsilon} \otimes c_{1,\ldots,N}$.
\end{enumerate}
One can always choose $N \leq 3$.

\item $H(\rho)\leq H(\sigma)$
\end{enumerate}
\end{thm}
\emph{Proof:} 
When Hamiltonians are trivial, thermal operations take the form \mbox{$\mathcal{E}(\rho) = \tr{2}{U(\rho \otimes \eta) U^{\dag}}$},
where $\eta$ is a maximally mixed state of any dimension and $U$ is an arbitrary unitary. However, this is the definition of a noisy operation~\cite{horodecki2003reversible, gour2013resource}. Furthermore, the existence of a noisy operation (mapping a given initial state arbitrarily close to a given target state) is equivalent to the majorization condition, so we conclude that $1 \Leftrightarrow 1'$, where
\begin{enumerate}
\item[1'.] For every $\epsilon >0$
\begin{enumerate}
\item  $\Vert \sigma - \sigma_{\epsilon}\Vert<\epsilon$,
\item  $\rho \otimes c_{1} \otimes \cdots \otimes c_N \succ \sigma_{\epsilon} \otimes c_{1,\ldots,N}$.
\end{enumerate}
\end{enumerate}
Then $1'\Rightarrow 2$. In fact, Schur-concavity and subadditivity of $H$ imply $H(\rho) \leq H(\sigma_{\epsilon})$ for every $\epsilon >0$. Then, by continuity, taking $\epsilon \rightarrow 0$ gives $H(\rho) \leq H(\sigma)$. Conversely, given $H(\rho) \leq H(\sigma)$, for any $0< \delta <1$ we can define the state $\sigma_{\delta} := (1-\delta)\sigma + \delta \mathbb{I}/d$, where $d$ is here the dimension of the Hilbert space of the system that carries the state $\rho$. Clearly for any fixed $\epsilon$ there exists $\delta$ small enough such that \mbox{$||\sigma - \sigma_{\delta}|| \leq \epsilon$}. Since $H$ is strictly concave, we get $H(\rho) \leq H(\sigma) < H(\sigma_{\delta})$. Also, by construction, ${\rm rank}(\rho)\leq {\rm rank}(\sigma_{\delta})$. Hence Theorem~\ref{Thm1Main} implies $1'$. This concludes the proof that $1' \Leftrightarrow 2$ and hence $1 \Leftrightarrow 2$.
\qed

We now introduce the tools necessary to extend this result to non-trivial Hamiltonians.

\subsubsection{The embedding map}

Given a vector of positive integers $\dd=\left\{ d_{1},\ldots,d_{n}\right\} $,
we can define the embedding map $\Gamma_{\dd}$, acting on a $n$-dimensional probability distribution
$p$ as follows \cite{brandao2013second}:
\[
\Gamma_{\dd}\left(p\right):=\bigoplus_{i=1}^{n}p_{i}\eta_{i},
\]
where the $\eta_{i}$ are $d_{i}$-dimensional uniform distributions. The embedding map allows us to move from a ``canonical'' theory governed by free energies $\{F_{\alpha}\}$ to a ``microcanonical'' theory governed by entropies $\{H_{\alpha}\}$ \cite{lostaglio2015description}. This is because if we denote by $\gamma$ a thermal Gibbs state with respect to the Hamiltonian $\HH$, we can find $\dd$ such that \mbox{$\Gamma_{\dd}(\gamma) = \eta$}, where $\eta$ is a uniform distribution of dimension \mbox{$D= \sum_i d_i$} (thus, it is the thermal state of a $D$-dimensional system with trivial Hamiltonian). Such a map exists whenever all entries of $\gamma$ are rational numbers, but we will drop this restriction later. 

We can then embed any probability distribution from $\mathbb{R}^n$ in $\mathbb{R}^D$ via $p \mapsto \Gamma_{\dd}(p)$. If $\rho$ is a quantum state block-diagonal in energy -- with occupations given by $p$ and corresponding Gibbs state $\gamma$ -- then the free energies of $\rho$ can be computed from the entropies of the embedded distribution (Lemma 9 in \cite{brandao2013second} and Eq.~\eqref{eq:trivial}):
\be
\label{eq:embedding}
F_\alpha(\rho)-F_\alpha(\gamma)=k T\left(\strut \log D -H_\alpha(\Gamma_{\dd}(p))\right).
\ee

\subsubsection{Result for non-trivial Hamiltonians}
We now rewrite Theorem~1 in order to make all the claims mathematically precise. As before, $\Vert \cdot \Vert$ can be any norm.

\begin{thm}
\label{theorem2}
Consider a system with Hamiltonian $\HH_S$ and states $\rho$ and $\sigma$ block-diagonal in energy. The three following statements are equivalent:

\begin{enumerate}
\item \label{claim1} For every $\epsilon >0$ there exist $N$ auxiliary systems with some Hamiltonians, in
a state $c_{1,\ldots,N}$ with marginals $c_1$,...,$c_N$, and a thermal operation $\mathcal{E}_{\epsilon}$ such that
\begin{enumerate}
\item  $\Vert \sigma - \sigma_{\epsilon}\Vert<\epsilon$,
\item  $\mathcal{E}_{\epsilon}(\rho \otimes c_{1} \otimes \cdots \otimes c_N)= \sigma_{\epsilon} \otimes c_{1,\ldots,N}$.
\end{enumerate}
\item \label{claim2} For every $\epsilon >0$ there exists
a state $c_{1,\ldots,N}$ on $N$ auxiliary systems with trivial Hamiltonians such that for every $\alpha \in \R$,
\[
F_{\alpha}(\rho) - kT \sum_i H_{\alpha} (c_i) \geq F_{\alpha}(\sigma_{\epsilon}) - kTH_{\alpha}(c_{1,\ldots,N}).
\]
\item \label{claim3} $F(\rho)\geq F(\sigma)$
\end{enumerate}
Notice that $c_{1,\ldots,N}$ will in general depend on $\epsilon$. It will follow from the proof that one can always choose $N = 3$ and trivial Hamiltonians for
the auxiliary systems in \ref{claim1}.
\end{thm}

\emph{Proof}:
For the sake of the proof, we introduce two further statements that will turn out to be equivalent:
\begin{enumerate}
\setcounter{enumi}{3}
\item \label{claim4} Identical to statement~\ref{claim1}, but with the additional requirement that the auxiliary systems have trivial Hamiltonians.
\item \label{claim5} For every $\epsilon >0$ there exist $N$ auxiliary systems with some Hamiltonians
and a state $c_{1,\ldots,N}$ on them such that for every $\alpha \in \R$,
\be
F_{\alpha}(\rho) + \sum_i F_{\alpha} (c_i) \geq F_{\alpha}(\sigma_{\epsilon}) +F_{\alpha}(c_{1,\ldots,N}).
\label{eq:freeenergyconditions}
\ee
\end{enumerate}
Clearly the implications $\ref{claim4}\Rightarrow\ref{claim1}$ and $\ref{claim2}\Rightarrow\ref{claim5}$ are trivially true.

$\ref{claim1} \Rightarrow \ref{claim5}$:  This follows immediately from the monotonicity of the $\alpha$-free energies $F_{\alpha}$ under thermal operations, Eq.~\eqref{eq:monotonicity},
and their additivity on tensor products. The same argument proves $\ref{claim4}\Rightarrow\ref{claim2}$.

\bigskip

$\ref{claim5} \Rightarrow \ref{claim3}$:  Take $\alpha=1$ in Eq.~\eqref{eq:freeenergyconditions}. Since the total Hamiltonian on the $N$ auxiliary systems is by definition
the sum of the local Hamiltonians, and due to the subadditivity of the Shannon entropy, we obtain
\[
   F(c_{1,\ldots,N})= \sum_{i=1}^N \langle \HH_i\rangle - k T \, H(c_{1,\ldots,N})
   \geq \sum_{i=1}^N \langle \HH_i\rangle - k T\sum_{i=1}^N H(c_i)
   =\sum_{i=1}^N F(c_i).
\]
Thus~(\ref{eq:freeenergyconditions}) implies $F(\rho)\geq F(\sigma_\epsilon)$. Since this is true for all $\epsilon>0$, and $F$ is continuous, we also obtain $F(\rho)\geq F(\sigma)$.

\bigskip

$\ref{claim3} \Rightarrow \ref{claim4}$: The proof is based on the techniques developed in \cite{brandao2013second}. Let us define by $p$ and $q$ the eigenvalues of $\rho$ and $\sigma$.
Consider the thermal state $\gamma_S$ corresponding to the Hamiltonian $\HH_S$. We may assume that $q\neq\gamma_S$, otherwise~\ref{claim4} follows trivially,
with $N=0$ and $\mathcal{E}_\epsilon$ the map that prepares the free Gibbs state.

First we consider the case that all components of $\gamma_S$ are rational numbers. By assumption there exists then $\dd = \{d_1,\ldots,d_n\}$ such that
\begin{eqnarray*}
\gamma_S=\left\{ \frac{d_{1}}{D},\frac{d_{2}}{D},\ldots,\frac{d_{n}}{D}\right\},  \quad \sum_{i=1}^{n}d_{i}=D.
\end{eqnarray*}
By definition, $\Gamma_{\dd}(\gamma_S) = \eta$, where $\eta$ is a $D$-dimensional uniform distribution. We can introduce an auxiliary probability distribution \mbox{$q_{\delta}=\left(1-\delta\right)q+\delta\gamma_S$}
(with \mbox{$0<\delta< 1$}) such that \mbox{$\left\Vert q-q_{\delta}\right\Vert \leq\delta$}. Since Shannon entropy is strictly concave, $F$ is strictly convex, and so
\[
F\left(q_{\delta}\right)=F\left(\left(1-\delta\right)q+\delta\gamma_S \right)<F(q).
\]
This implies
\begin{equation}
F\left(p\right)\geq F\left(q\right)>F\left(q_{\delta}\right).\label{free_energy}
\end{equation}
From Eq.~\eqref{eq:embedding}, it follows that $H\left(\Gamma_{\dd}\left(p\right)\right)<H\left(\Gamma_{\dd}\left(q_{\delta}\right)\right)$.
Furthermore ${\rm rank}\left(p\right)\leq {\rm rank}\left(q_{\delta}\right)$
by construction, and this implies ${\rm rank}\left(\Gamma_{\dd}\left(p\right)\right)\leq {\rm rank}\left(\Gamma_{\dd}\left(q_{\delta}\right)\right)$. Using Theorem~\ref{Thm1Main}, this is equivalent to the existence of an auxiliary system $c_{1,\ldots,N}$ with marginals $c_1$,\ldots,$c_N$ and a noisy map $\Lambda$ such that

\begin{eqnarray*}
\Lambda\left(\Gamma_{\dd}\left(p\right)\otimes c_{1}\otimes\cdots\otimes c_{N}\right) & = & \Gamma_{\dd}\left(q_{\delta}\right)\otimes c_{1,\ldots,N},\\
\Lambda\left(\eta\otimes\eta_{1}\otimes\cdots\otimes\eta_{N}\right) & = & \eta\otimes\eta_{1}\otimes\cdots\otimes\eta_{N}.
\end{eqnarray*}

If we consider the map $\Lambda_{th}=\left(\Gamma^{-1}_{\dd}\otimes\mathbb{I}\right)\circ\Lambda\circ\left(\Gamma_{\dd}\otimes\mathbb{I}\right)$ (with the identity acting on the space of the auxiliary systems),
it is easy to check that it is Gibbs-preserving, and it also maps $p$ to $q$ while correlating the auxiliary systems:

\begin{eqnarray*}
\Lambda_{th}\left(\gamma_S\otimes\eta_{1}\otimes\cdots\otimes\eta_{k}\right) & = & \gamma_S\otimes\eta_{1}\otimes\cdots\otimes\eta_{k},\\
\Lambda_{th}\left(p\otimes c_{1}\otimes\cdots\otimes c_{k}\right) & = & q_{\delta}\otimes c_{1,\ldots,N}.
\end{eqnarray*}
The existence of a Gibbs-preserving map is equivalent to the existence of a thermal map in the ``semiclassical'' case of block-diagonal states, because
both are characterized by the same set of constraints \cite{faist2015gibbs}.
Using this equivalence, we obtain \ref{claim4} with $\varepsilon=\delta$.

Now let us consider the case of $\gamma_S$ with general real entries. This extension follows the proof of Theorem 17 in~\cite{brandao2013second}, but we sketch it here for convenience. From Lemma 15 of \cite{brandao2013second}, we can construct
a channel $E$ that maps $\gamma_S$ into an $\varepsilon$-close distribution
$\tilde{\gamma}$ with rational entries, while not perturbing $p$ too much; more precisely, we can define a stochastic map $E$ such
that 
\begin{eqnarray*}
E\left(\gamma_S\right)=\tilde{\gamma}_S;  \quad  \quad \left\Vert \gamma_S-\tilde{\gamma}_S\right\Vert \leq\varepsilon,\\
\\
 \left\Vert p-E\left(p\right)\right\Vert \leq \mathcal{O}\left(\sqrt{\varepsilon}\right) \textrm{  for all } p.
\end{eqnarray*}
If $\varepsilon$ is small enough, is it possible to follow \cite{brandao2013second}
and show that 
\[
S\left(E\left(p\right)\Vert\tilde{\gamma}_S\right)>S\left(E\left(q_{\delta}\right)\Vert\tilde{\gamma}_S\right).
\]
This is equivalent to the condition $F(E(p))>F(E(q_{\delta}))$, as long as we remember that now $F$ is defined w.r.t. the thermal state $\tilde{\gamma}_S$. Since
$\tilde{\gamma}_S$ has rational entries we use the first part of the proof and find a thermal map $\tilde{\Lambda}_{th}$:
\begin{eqnarray*}
\tilde{\Lambda}_{th}\left(\tilde{\gamma}_S\otimes\eta_{1}\otimes\cdots\otimes\eta_{k}\right) & = & \tilde{\gamma}_S\otimes\eta_{1}\otimes\cdots\otimes\eta_{k},\\
\tilde{\Lambda}_{th}\left(E(p)\otimes c_{1}\otimes\cdots\otimes c_{k}\right) & = & E(q_{\delta})\otimes c_{1,\ldots,N}.
\end{eqnarray*}
Using again Lemma 15 of \cite{brandao2013second}, we can find a stochastic map $E'$ mapping $\tilde{\gamma}_S$ into $\gamma_S$ without perturbing other probability distributions a lot. If we consider the map $\Lambda_{th}=\left(E' \otimes\mathbb{I}\right)\circ\tilde{\Lambda}_{th}\circ\left(E\otimes\mathbb{I}\right)$
(where $\mathbb{I}$ acts only on the auxiliary systems), one finds $\Lambda_{th}$ satisfying
\begin{eqnarray*}
\Lambda_{th}\left(\gamma_S\otimes\eta_{1}\otimes\cdots\otimes\eta_{k}\right) & = & \gamma_S\otimes\eta_{1}\otimes\cdots\otimes\eta_{k},\\
\Lambda_{th}\left(p\otimes c_{1}\otimes\cdots\otimes c_{k}\right) & = & E'\circ E (q_{\delta})\otimes c_{1,\ldots,N}.
\end{eqnarray*}
Furthermore $\left\Vert E'\circ E\left(q_{\delta}\right)-q\right\Vert \leq\delta+O\left(\sqrt{\varepsilon}\right)$
and, since $\delta$ and $\varepsilon$ can be chosen arbitrarily small, we recover the first part of the proof and also obtain~\ref{claim4}.
\qed

\subsection{Comparison with other work extraction models}

There is a subtle but important point that needs to be clarified when comparing our framework with other approaches to work extraction. In the context of single-shot thermodynamics, recent results fell into two categories: one approach is to demand deterministic work extraction, studied for example in \cite{brandao2013second}, where no probability of error is allowed. They require a deterministic transformations of the work bit from its ground state to the excited state \mbox{$\ket{0} \rightarrow \ket{1}$} (this corresponds to $\epsilon =0$ in Eq.~(7)) and show that one can extract $w= F_0(\rho)$, \emph{provided} that an arbitrarily small amount of work is consumed in the process. Specifically, they allow the use of a system with trivial Hamiltonian and initially in a state $\ket{0}$, that is returned arbitrarily close to its initial state at the end of the process. Another approach is the so-called almost-deterministic work extraction, followed for example in \cite{aberg2013truly}, where a fixed probability of failure is allowed. Looking at Eq.~(1) one may be tempted to conclude that we are looking at the latter case.

However, our main result (Theorem~1) is about transformations that can be performed with \emph{arbitrarily small} error probability $\epsilon$. The same arbitrarily small error probability appears in the work extraction protocol of \cite{brandao2013second} as well, however it goes in the pure state with trivial Hamiltonian that they allow to introduce. Hence the difference between the two approaches is technical and not physical. Physically speaking, being able to ensure an arbitrarily small error probability $\epsilon$ in $\chi_{\epsilon}(w)$ is operationally indistinguishable from a deterministic protocol. For this reason, we preferred to avoid invoking the use of a pure state as an extra resource.

\subsection{Arbitrarily small amounts of correlation are needed}
\label{smallcorrelations}
Let us now present a simple argument showing that one needs to generate only an arbitrarily small amount of correlations among the auxiliary systems to induce any transformation between two states $\rho$ and $\sigma$ with $F(\rho)\geq F(\sigma)$. For this purpose, we introduce an extra system in a thermal state $\gamma_S$, and we consider the transformation $\rho \otimes \gamma_S \rightarrow \sigma \otimes \tau$, 
where $\tau$ is chosen such that $F(\tau) = F(\rho) + F(\gamma_S) - F(\sigma) $. By construction the free energy $F$ of the left-hand-side equals the free energy of the right-hand-side, so Theorem~1 applies. Hence we can find $c_{1,..,N}$ with marginals $c_1$, \ldots,$c_N$ such that
\[
\rho \otimes \gamma_S \otimes c_1 \otimes \ldots \otimes c_N \rightarrow (\sigma \otimes \tau)_{\epsilon} \otimes c_{1,\ldots,N},
\]
where $(\sigma \otimes \tau)_{\epsilon}$ is a state $\epsilon$-close to $\sigma \otimes \tau$ {in trace distance, and $\epsilon>0$ can be chosen arbitrarily small. We can discard the extra system and obtain $\sigma_{\epsilon}$ $\epsilon$-close to $\sigma$ due to the monotonicity of the trace distance under partial traces. Computing $F$ on both sides of the previous equation, using the monotonicity of $F$ and its additivity under tensor products, we obtain
\begin{eqnarray*}
   I(c_{1,\ldots,N})&\leq& \frac{F(\rho)+F(\gamma_S)-F\left(\strut (\sigma\otimes\tau)_\epsilon\right)}{kT}
   = \frac{F(\sigma\otimes\tau)-F\left(\strut (\sigma\otimes\tau)_\epsilon\right)}{kT}.
\end{eqnarray*}
From the continuity of $F$, it follows that
\[
I(c_{1,..,N}) \stackrel{\epsilon\rightarrow 0}{\longrightarrow} 0.
\]
We conclude that by choosing $\epsilon$ arbitrarily small, we can make the total correlations $I$ arbitrarily weak.

\subsection{Dimension of the auxiliary systems}

Let $\rho$ and $\sigma$ be two quantum states block-diagonal in energy such that 
\[
F(\rho)=F(\sigma).
\]
Suppose that some of the free energy conditions of Eq.~(\ref{eq:monotonicity}) are not satisfied (otherwise, a transformation that maps $\rho$ to a state arbitrarily
close to $\sigma$ could trivially be performed without creating any correlation between the auxiliary systems, cf.\ Appendix~\ref{operations}).
Then we know that there exists some $\alpha>0$, $\alpha \neq 1$, s.t. 
\be
\label{eq:nontrivial}
F_{\alpha}(\rho) < F_{\alpha} (\sigma).
\ee
According to Theorem~\ref{theorem2}, for every $\epsilon >0$ there exists a thermal operation $\mathcal{E}_\epsilon$ and $c^{\epsilon}_{1,\ldots,n}$ such that
\be
\label{eq:ccto}
\mathcal{E}_\epsilon(\rho \otimes c^{\epsilon}_1 \otimes \ldots \otimes c^{\epsilon}_n) = \sigma_{\epsilon} \otimes c^{\epsilon}_{1,\ldots,n}, \quad \|\sigma_{\epsilon} - \sigma\| \leq \epsilon.
\ee
It follows immediately from the result of Appendix~F that this implies 
\[
   I(c_{1,\ldots,n}^\epsilon)=S(c^{\epsilon}_{1,\ldots,n}\|c^{\epsilon}_1 \otimes \ldots \otimes c^{\epsilon}_n) \stackrel{\epsilon \rightarrow 0}{\longrightarrow} 0.
\]
Due to Pinsker's inequality this also implies 
\be
\label{eq:normtozero}
\|c^{\epsilon}_{1,\ldots,n} - c^{\epsilon}_1 \otimes \ldots \otimes c^{\epsilon}_n\|  \stackrel{\epsilon \rightarrow 0}{\longrightarrow} 0.
\ee
At the same time, Eq.~\eqref{eq:ccto} and Eq.~\eqref{eq:entropicconditions} imply that \emph{all} free energy second laws are satisfied for all $\epsilon >0$. In particular then for every $\epsilon > 0$,
\[
F_{\alpha}(\rho) - F_{\alpha} (\sigma_\epsilon) \geq kT [H_{\alpha}(c^{\epsilon}_1 \otimes \ldots \otimes c^{\epsilon}_n) - H_{\alpha}(c^{\epsilon}_{1,\ldots,n})].
\]
By Eq.~\eqref{eq:nontrivial} and by continuity, the left-hand side turns to $F_\alpha(\rho)-F_\alpha(\sigma)<0$ for $\epsilon\to 0$. Thus
\begin{equation}
   \limsup_{\epsilon\to 0}\left( H_{\alpha}(c^{\epsilon}_1 \otimes \ldots \otimes c^{\epsilon}_n)   - H_{\alpha}(c^{\epsilon}_{1,\ldots,n})\right) <0.
   \label{eqRenyiDifference}
\end{equation}
Assume for now that $\alpha>1$. Then, using \mbox{$H_{\alpha}(p)= \frac{1}{1-\alpha} \log \|p\|^\alpha_{\alpha}$}, we obtain
\[
\liminf_{\epsilon\to 0}\frac{\|c^{\epsilon}_1 \otimes \ldots \otimes c^{\epsilon}_n\|_{\alpha}}{\|c^{\epsilon}_{1,\ldots,n}\|_{\alpha}} > 1.
\]
Using the triangle inequality on the numerator of the above expression,
\[
\frac{\|c^{\epsilon}_1 \otimes \ldots \otimes c^{\epsilon}_n\|_{\alpha}}{\|c^{\epsilon}_{1,\ldots,n}\|_{\alpha}}\leq 1 + \frac{\|c^{\epsilon}_{1,\ldots,n} - c^{\epsilon}_1 \otimes \ldots \otimes c^{\epsilon}_n)\|_{\alpha}}{\|c^{\epsilon}_{1,\ldots,n}\|_{\alpha}},
\]
that implies
\be
\label{eq:liminf}
\liminf_{\epsilon \rightarrow 0} \frac{\|c^{\epsilon}_{1,\ldots,n} - c^{\epsilon}_1 \otimes \ldots \otimes c^{\epsilon}_n)\|_{\alpha}}{\|c^{\epsilon}_{1,\ldots,n}\|_{\alpha}} >0.
\ee
However from the equivalence of all norms in finite dimension and Eq.~\eqref{eq:normtozero},
\[
   \|c^{\epsilon}_{1,\ldots,n} - c^{\epsilon}_1 \otimes \ldots \otimes c^{\epsilon}_n\|_{\alpha} \stackrel{\epsilon \rightarrow 0}{\longrightarrow} 0.
\]
For this to be possible and the inequality \eqref{eq:liminf} to be satisfied we need
\be
\label{eq:alphanormtozero}
\limsup_{\epsilon \rightarrow 0} \|c^{\epsilon}_{1,\ldots,n}\|_\alpha=  0 = \lim_{\epsilon\to 0}\|c^{\epsilon}_{1,\ldots,n}\|_\alpha.
\ee
So we have a sequence of auxiliary systems satisfying the previous equation and the obvious normalization condition
\[
\|c^{\epsilon}_{1,\ldots,n}\|_{1} = 1 \quad \forall \epsilon>0.
\]
However, we now use an inequality for the $p$-norms:
\be
\label{eq:bounda}
1 = \|c^{\epsilon}_{1,\ldots,n}\|_{1} \leq n(\epsilon)^{1-1/\alpha} \|c^{\epsilon}_{1,\ldots,n}\|_{\alpha} \quad  \forall \epsilon>0,
\ee
where $n(\epsilon)$ is the product of the local dimensions of the $n$ auxiliary systems.
For Eq.~\eqref{eq:alphanormtozero} and Eq.~\eqref{eq:bounda} to be simultaneously satisfied, $n(\epsilon)$ must tend to infinity as $\epsilon \rightarrow 0$.

Now consider the case $0<\alpha<1$. We have the following inequality from~\cite[Theorem 4]{rastegin2013bounds}: set $T:=\frac 1 2 \|p-q\|_1$, then
for all probability distributions $p,q\in\R^d$,
\begin{equation}
   \left| H_\alpha(p)-H_\alpha(q)\right| \leq\frac{T^\alpha\left(\strut d(d-1)\right)^{1-\alpha}-\alpha T}{1-\alpha}.
   \label{eqRastegin}
\end{equation}
(Actually, \cite{rastegin2013bounds} shows this for the Tsallis entropy, but via $|\log x-\log y|\leq |x-y|$ for $x,y\geq 1$ one can easily see that the bound carries
over to R\'enyi entropy.)
Suppose the dimension of the catalysts was bounded, i.e.\ $n(\epsilon)\leq d$ for all $\epsilon>0$, with $d\in\mathbb{N}$ some fixed number.
Then Eq.~\eqref{eqRastegin} and Eq.~\eqref{eq:normtozero} would imply that
\mbox{$\lim_{\epsilon\to 0}H_{\alpha}(c^{\epsilon}_1 \otimes \ldots \otimes c^{\epsilon}_n) - H_{\alpha}(c^{\epsilon}_{1,\ldots,n})=0$}, in contradiction to Eq.~\eqref{eqRenyiDifference}.

That is, in all cases, we conclude that the dimension of the catalysts has to grow unboundedly if the accuracy $\epsilon$ tends to zero.

\end{document}